\documentclass[preprint,pre,aps,amsfonts,amsmath,showpacs,superscriptaddress,nofootinbib]{revtex4}
\usepackage{amsmath,amssymb,amscd}
\usepackage{graphicx}
\usepackage{bm}

\begin{document}
%

\title{First Gap Statistics of Long Random Walks with Bounded Jumps}
\author{Philippe Mounaix}
\email{philippe.mounaix@cpht.polytechnique.fr}
\affiliation{Centre de Physique Th\'eorique, Ecole
Polytechnique, CNRS, Universit\'e Paris-Saclay, F-91128 Palaiseau, France.}
\author{Gr\'egory Schehr}
\email{gregory.schehr@lptms.u-psud.fr}
\affiliation{LPTMS, CNRS, Univ. Paris-Sud, Universit\'e Paris-Saclay, 91405 Orsay, France.}
\date{\today}
\begin{abstract}
We study one-dimensional discrete as well as continuous time random walks, either with a fixed number of steps (for discrete time)
$n$ or on a fixed time interval $T$ (for continuous time). In both cases, we focus on symmetric probability distribution functions (PDF) of jumps with a finite support $[-g_{max}, g_{max}]$. For continuous time random walks (CTRWs), the waiting time $\tau$ between two consecutive jumps is a random variable whose probability distribution (PDF) has a power law tail $\Psi(\tau) \propto \tau^{-1-\gamma}$, with $0<\gamma<1$. We obtain exact results for the joint statistics of the gap between the first two maximal positions of the random walk and the time elapsed between them. We show that for large $n$ (or large time $T$ for CTRW), this joint PDF reaches a stationary joint distribution which exhibits an interesting concentration effect in the sense that a gap close to its maximum possible value, $g\approx g_{max}$, is much more likely to be achieved by two successive jumps rather than by a long walk between the first two maxima. Our numerical simulations confirm this concentration effect.     
\end{abstract}
\pacs{05.40.Fb, 02.50.Cw}
\maketitle
%
%
\section{Introduction and summary of main results}\label{sec1}
Extreme value statistics (EVS) of random walks (RWs) -- and their continuous counterpart, Brownian motion -- has recently attracted much attention (for recent short reviews see \cite{review1, review2}). On the one hand, it has indeed been realized that RWs provide a very interesting laboratory to test and characterize the effects of strong correlations between the walker positions at different times on EVS quantitatively, about which very little is known (unlike the uncorrelated case of independent and identically distributed (i.i.d.) random variables \cite{Gumbel}). On the other hand, it was shown that EVS of RWs have several interesting applications in statistical physics (e.g. to disordered systems \cite{PLD}, fluctuating interfaces \cite{airy1, airy2, airy3} or $1/f^\alpha$ noise \cite{GMOR2007}), finance (e.g. in optimal portfolio strategy \cite{MB2008}), computer science (e.g. in data storage analysis \cite{FPV1998} or in the statistics of tree structures \cite{Majreview}) or even in random geometry in two-dimensions (e.g. for the convex hull of two-dimensional stochastic processes \cite{convex1, convex2, convex3} or in random convex geometry \cite{Sylvester}). Although in many cases the considered RWs consist of unbounded random jumps (e.g. Gaussian or $\alpha$-stable distributed jumps), RWs with bounded jumps have also interesting applications like, for instance, in a well studied packing problem in two dimensions where $n$ rectangles of variable sizes are packed in a semi-infinite strip of width one. In this case, it can be shown that the fluctuations of the height of the optimal packing are given by the maximum of a RW whose random jumps are bounded and uniformly distributed over the interval $[-1,+1]$ (see Refs. \cite{packing1, packing2, packing3}). More generally, lots of real world random walks are expected to have bounded jumps because of natural limitations inherent in the underlying physics and, in this respect, it is useful to get a good understanding of this case.

Although these applications concern the {\it global} maximum of the RW, one may also wonder about the statistics of the second, third, $\cdots$, more generally of the $k$-th maximum $M_k$ of a RW ($M_1$ being just the global maximum). This is known under the name of `order statistics' which was recently studied for RWs \cite{order_rw} as well as for more general stochastic processes, like branching Brownian motion \cite{orderBBM1,orderBBM2,orderBBM3}, or $1/f^\alpha$ noise \cite{order1overf}. Interestingly, these works demonstrated that the order statistics of strongly correlated variables have a very rich structure, much richer than their i.i.d. counterpart. More recently, we investigated the statistics of the first gap $g = M_1 - M_2$ as well as the time elapsed between the first two maxima, both for discrete time RWs \cite{MMS2013, MMS2014} and continuous time random walks (CTRWs) \cite{MMS2016}. Quite remarkably, we found that the behavior of the joint probability distribution function (PDF) of the gap and the time is very sensitive to the distribution of the jumps defining the RW. Our previous investigations covered a wide range of jumps distributions but ignored the important case of {\it bounded} distributions, which is the subject of the present work completing our earlier ones on the complementary case of unbounded jumps\ \cite{MMS2013,MMS2014,MMS2016}.   

This paper deals with the statistics of the gap and time interval between the first two maxima of long random walks with {\it bounded} jumps. One of the motivations for this work has been to investigate the `concentration' phenomenon observed and studied in\ \cite{MMS2014,MMS2016} for unbounded jumps with distributions going to zero fast enough at infinity (what we called `fast decreasing jump distribution'). For such fast decreasing distributions a large gap is more likely to be achieved over two successive jumps instead of over a longer walk between the first two maxima. As we will see in the following, (i) the conjecture made in\ \cite{MMS2014,MMS2016} that a similar concentration should exist for jump distributions with a bounded support was correct, and (ii) there are interesting, non trivial, differences between the bounded jump problem and the fast decreasing jump distribution problem of\ \cite{MMS2014,MMS2016}.

In this paper, we consider a random walk starting at the origin, $x_0 = 0$, and evolving according to
\begin{equation}\label{def_RW}
x_i = x_{i-1} + \eta_{i},
\end{equation}
where $x_i$ denotes the walker position between the $i$-th and the $(i+1)$-th jumps. The jumps $\eta_i$'s are i.i.d. random variables distributed following a symmetric, bounded and piecewise continuous distribution $f(\eta)$ with a bounded support $-g_{max}\le\eta\le g_{max}$ (with fixed $g_{max}>0$), the Fourier transform of which, $\hat{f}(k)=\int_{-g_{max}}^{g_{max}}f(\eta)\, \exp(ik\eta)\, d\eta$, has the small $k$ behavior
\begin{equation}\label{small_k}
\hat{f}(k)=1-\vert ak\vert^2 +o(\vert k\vert^2),
\end{equation}
with $a=\sigma/\sqrt{2}$, where $\sigma^2=\int_{-g_{max}}^{g_{max}}\eta^2f(\eta)\, d\eta$ is the variance of the jump distribution. Let $\tau_i$ denote the time interval between the $(i-1)$-th and the $i$-th jumps ($\tau_i>0$). In the case of discrete time random walks one simply has $\tau_i =1$, while for continuous time random walks the $\tau_i$'s are i.i.d. continuous random variables, independent of $\eta_i$, with PDF $\Psi(\tau)$ the Laplace transform of which, $\hat\Psi(q)=\int_0^{+\infty}\Psi(\tau)\, \exp(-q\tau)\, d\tau$, has the small $q$ behavior
\begin{equation}\label{def_gamma}
\hat\Psi(q)=1-(\tau_c q)^\gamma +o(\vert q\vert^\gamma),
\end{equation}
where $0<\gamma\le 1$ and $\tau_c >0$ is the characteristic time scale of the jumps. For $\gamma =1$, the mean time between two successive jumps, $\langle\tau\rangle\equiv\int_0^{+\infty}\tau\Psi(\tau)\, d\tau <+\infty$, exists and $\tau_c=\langle\tau\rangle$. For $0<\gamma <1$, the mean time between two successive jumps does not exist.

Our approach to the problem for long random walks is based on the result, proved in\ \cite{MMS2013,MMS2014}, that the joint PDF $p_n(g,l)$ of the gap $g$ and the number of jumps $l$ between the first two maxima of the sequence $\lbrace x_0, x_1, x_2, \cdots ,x_n\rbrace$ (i.e. a random walk of $n$ jumps) has a well defined limiting PDF as $n\rightarrow +\infty$. More specifically, one has
\begin{equation}\label{lim_PDF1}
\lim_{n\rightarrow +\infty}p_n(g,l)=p(g,l),
\end{equation}
where $p(g,l)$ is given by its generating function with respect to $l$,
\begin{equation}\label{def_GF}
\tilde{p}(g,s)=\sum_{l>0}p(g,l)s^l
=I_1(g,s)I_2(g),
\end{equation}
with
\begin{equation}\label{def_I1_I2}
\begin{array}{l}
I_1(g,s)=s\int_0^{g_{max}-g}u(x,s) f(g+x)\, dx, \\
I_2(g)=\int_0^{g_{max}-g}h(x,1) f(g+x)\, dx,
\end{array}
\end{equation}
where the upper bound at $x=g_{max}-g$ results from the bounded support of $f(\eta)$, $-g_{max}\le\eta\le g_{max}$ (with fixed $g_{max}>0$). In Eq. (\ref{def_I1_I2}) the functions $u(x,s)$ and $h(x,s)$ are defined by their Laplace transforms with respect to $x$,
\begin{equation}\label{def_u_h}
\begin{array}{l}
\int_0^{+\infty}u(x,s){\rm e}^{-\lambda x}dx=\phi(\lambda ,s), \\
\int_0^{+\infty}h(x,s){\rm e}^{-\lambda x}dx=\phi(\lambda ,s)/\lambda ,
\end{array}
\end{equation}
with
\begin{equation}\label{def_Phi}
\phi(\lambda ,s)=\exp\left(-\frac{\lambda}{\pi}
\int_0^{+\infty}\frac{\ln\lbrack 1-s\hat{f}(k)\rbrack}{k^2+\lambda^2}\, dk\right) .
\end{equation}
A limit similar to\ (\ref{lim_PDF1}) holds for continuous time random walks as well\ \cite{MMS2016}. Namely, replacing $p_n(g,l)$ with $p_n(g,t)$, the joint PDF of the gap $g$ and the time interval $t$ between the first two maxima of the sequence $\lbrace x_0, x_1, x_2, \cdots ,x_n\rbrace$, it is not difficult to show that $p_n(g,t)$ has a well defined limiting PDF too as $n\rightarrow +\infty$. One finds (see\ \cite{MMS2016} for details),
\begin{equation}\label{lim_PDF2}
\lim_{n\rightarrow +\infty}p_n(g,t)=p(g,t),
\end{equation}
where $p(g,t)$ is given by its Laplace transform with respect to $t$ in terms of $\tilde{p}$ and $\hat{\Psi}$,
\begin{equation}\label{def_LT}
\hat p(g,q)=\int_0^{+\infty}p(g,t)\, \exp(-qt)\, dt=
\tilde p(g,\hat\Psi(q)).
\end{equation}
Equations\ (\ref{def_GF}) and\ (\ref{def_LT}) are the starting points of our study in the cases of discrete and continuous time, respectively.

Before entering the details of the calculations, it is useful to summarize our main results. We first show that for a discrete time random walk, $l$ concentrates onto $l=\pm 1$ as $g\rightarrow g_{max}$, the largest possible gap, in the sense that for all $l\ne\pm 1$, $p(g,\pm 1)/p(g,l)\rightarrow +\infty$ as $g\rightarrow g_{max}$. In the very wide class of jump distributions $f(\eta)$ behaving algebraically as $\eta\rightarrow g_{max}$ (or discontinuous to zero at $\eta = g_{max}$, see the remark below Eq.\ (\ref{eq2.10})), we find the universal behavior $p(g,\pm 1)/p(g,l)\sim (g_{max}-g)^{-1}$ as $g\rightarrow g_{max}$. This result is clearly confirmed by our numerical simulations (see Fig. \ref{Fig:concent_RW}). In the case of continuous time random walks, we perform a detailed analysis of the different asymptotic behaviors of the joint PDF $p(g,t)$ in the plane $(g,t)$. (Since $p(g,-t)=p(g,t)$, we take $t>0$ without loss of generality). The main result of this study is the existence of a scaling form for $p(g,t)$ when $0<\gamma <1$ and for algebraic $f(\eta)$ near $\eta =g_{max}$. More specifically, we show that there is a scaling regime $(g_{max}-g)^{-1},\ t \gg 1$ with fixed $(g_{max}-g)t^{\gamma/2}$, in which $p(g,t)$ takes the scaling form
\begin{eqnarray}\label{scaling1a}
p(g,t)\sim
\frac{a c^2}{\tau_c}
\, \frac{(g_{max}-g)^{3+2(\alpha +1/\gamma)}}{\lbrack a(\alpha +1)\rbrack^{3+2/\gamma}}
\, K\left(\frac{g_{max}-g}{a(\alpha +1)}\left(\frac{t}{\tau_c}\right)^{\gamma/2}\right)&& \nonumber \\
(g\stackrel{f.b.}{\longrightarrow} g_{max}\ {\rm and}\ t\rightarrow +\infty),&&
\end{eqnarray}
where
\begin{equation}\label{scaling1b}
K(y)=\frac{1}{y^{1+2/\gamma}}\, \left(\frac{\mathcal{D}_{\rm I}}{y}+\mathcal{D}_{\rm II}\right) ,
\end{equation}
with the asymptotic behaviors
\begin{equation}\label{scaling1c}
K(y)\sim\left\lbrace
\begin{array}{ll}
\mathcal{D}_{\rm I} \, y^{-2(1+1/\gamma)}&(y\rightarrow 0), \\
\mathcal{D}_{\rm II} \, y^{-1-2/\gamma}&(y\rightarrow +\infty).
\end{array}\right.
\end{equation}
The amplitudes $\mathcal{D}_{\rm I}$ and $\mathcal{D}_{\rm II}$ are given in Eqs.\ (\ref{def_D1}) and\ (\ref{def_D2}), respectively. Physically, the switch from the first to the second behavior\ (\ref{scaling1c}) around $(t/\tau_c)^{\gamma/2}\sim a(g_{max}-g)^{-1}$ corresponds to the cross-over from a `concentration' -- or `one-jump' --  regime (for $(t/\tau_c)^{\gamma/2}< a(g_{max}-g)^{-1}$) where the walker gets stuck for a long time $t$ at the second maximum and then jumps directly to the first maximum, to a `many-jumps' regime (for $(t/\tau_c)^{\gamma/2}> a(g_{max}-g)^{-1}$) where she/he travels a long walk of total duration $t$ (with many steps) between the second and the first maxima. For $\gamma =1$, or non algebraic $f(\eta)$ near $\eta =g_{max}$, there is no scaling form and\ (\ref{scaling1a}) is replaced with the uniform expression\ (\ref{eq3.8}) from which it is possible to determine the domains in the $(t,g)$ plane corresponding to `one-jump' --  and `many-jumps' regimes in this case too.

The outline of the paper is as follows. Section\ \ref{sec2} deals with the concentration of $l$ onto $l=\pm 1$ in the case of discrete time random walks. The effects of this concentration on continuous time random walks, especially the `concentration' and `many-jumps' asymptotic regimes of $p(g,t)$ for large $t$ and small $g_{max}-g$, are studied in Section\ \ref{sec3}. Finally, Section\ \ref{sec4} is devoted to the comparison of our analytical results with numerical simulations.
%
%
%
\section{Discrete time random walks}\label{sec2}
\subsection{Concentration of $\bm{l}$ onto $\bm{l=\pm 1}$ as $\bm{g\rightarrow g_{max}}$}\label{sec2.1}
As mentioned at the end of Sec. 4.3 in Ref.\ \cite{MMS2014}, if the jump distribution $f(\eta)$ has a bounded support, $-g_{max}\le\eta\le g_{max}$ (with fixed $g_{max}>0$), a concentration of $l$ onto $l=\pm 1$ is expected to occur in the limit $g\rightarrow g_{max}$, the largest possible value of the gap. Practically, this means that a large gap, close to its largest possible value, is expected to be mainly due to configurations with adjacent first and second maxima. To prove that this is actually the case we must determine the behavior of $p(g,l)$ for $g<g_{max}$ close to $g_{max}$. To this end we need the small $x$ behavior of the functions $u(x,s)$ and $h(x,s)$ defined in\ (\ref{def_u_h}), which is obtained from the large $\lambda$ behavior of $\phi(\lambda ,s)$ given in\ (\ref{def_Phi}). One has
\begin{equation}\label{eq2.1}
\phi(\lambda ,s)\sim 1-\frac{1}{\pi\lambda}\int_0^{+\infty}\ln\lbrack 1-s\hat{f}(k)\rbrack\, dk\ \ \ \ (\lambda\rightarrow +\infty),
\end{equation}
from which it follows
\begin{eqnarray}\label{eq2.2}
u(x,s)&=&\frac{1}{2i\pi}\int_{\mathcal{L}}\phi(\lambda ,s)\, {\rm e}^{\lambda x}d\lambda \nonumber \\
&\sim& \delta(x-0^+)-\frac{1}{\pi}\int_0^{+\infty}\ln\lbrack 1-s\hat{f}(k)\rbrack\, dk\ \ \ \ (x\rightarrow 0),
\end{eqnarray}
and
\begin{equation}\label{eq2.3}
h(x,s)=\frac{1}{2i\pi}\int_{\mathcal{L}}\phi(\lambda ,s)\, {\rm e}^{\lambda x}\frac{d\lambda}{\lambda}
\sim 1\ \ \ \ (x\rightarrow 0).
\end{equation}
Thus, by injecting\ (\ref{eq2.2}) and\ (\ref{eq2.3}) into the definition\ (\ref{def_I1_I2}) of $I_1(g,s)$ and $I_2(g)$ one gets
\begin{eqnarray}\label{eq2.4}
&&I_1(g,s)=s\int_0^{g_{max}-g}u(x,s)f(g+x)\, dx \\
&&\sim sf(g)-\int_0^{g_{max}-g}f(g+x)\, dx\, 
\frac{s}{\pi}\int_0^{+\infty}\ln\lbrack 1-s\hat{f}(k)\rbrack\, dk
\ \ \ \ (g\stackrel{f.b.}{\longrightarrow} g_{max}),\nonumber
\end{eqnarray}
and
\begin{eqnarray}\label{eq2.5}
I_2(g)&=&\int_0^{g_{max}-g}h(x,1)f(g+x)\, dx \nonumber \\
&\sim&\int_0^{g_{max}-g}f(g+x)\, dx
\ \ \ \ (g\stackrel{f.b.}{\longrightarrow} g_{max}),
\end{eqnarray}
where $\stackrel{f.b.}{\longrightarrow}$ means `{\it tends to, from below}'. Equations\ (\ref{def_GF}),\ (\ref{eq2.4}), and (\ref{eq2.5}) give
\begin{eqnarray}\label{eq2.6}
&&\tilde{p}(g,s)=I_1(g,s)I_2(g)\sim s\int_g^{g_{max}}f(x)\, dx \\
&&\times\left(f(g)-\frac{1}{\pi}\int_g^{g_{max}}f(x)\, dx\, 
\int_0^{+\infty}\ln\lbrack 1-s\hat{f}(k)\rbrack\, dk\right)
\ \ \ \ (g\stackrel{f.b.}{\longrightarrow} g_{max}),\nonumber
\end{eqnarray}
where we have made the change of variable $g+x\rightarrow x$ in the integrals over $x$ in\ (\ref{eq2.4}) and\ (\ref{eq2.5}).The joint PDF $p(g,l)$ is then readily obtained by expanding the logarithm in power series of $s$. One finds
\begin{equation}\label{eq2.7}
p(g,l)\sim\left\lbrace
\begin{array}{lr}
f(g)\int_g^{g_{max}}f(x)\, dx&l=\pm 1,\\
\frac{p(0,\vert l\vert -1\vert 0,0)}{\vert l\vert -1}\left(\int_g^{g_{max}}f(x)\, dx\right)^2&\vert l\vert\ge 2,
\end{array}\right.
\ \ \ \ (g\stackrel{f.b.}{\longrightarrow} g_{max}),
\end{equation}
where we have rewritten the integrals over $k$ as
\begin{equation}\label{eq2.8}
\frac{1}{\pi}\int_0^{+\infty}\hat{f}(k)^n dk=
\frac{1}{2\pi}\int_{-\infty}^{+\infty}\hat{f}(k)^n dk=
p(0,n\vert 0,0).
\end{equation}
From\ (\ref{eq2.7}) it follows immediately that, for all $l$ with $\vert l\vert\ge 2$,
\begin{equation}\label{eq2.9}
\frac{p(g,\pm 1)}{p(g,l)}\sim\frac{\vert l\vert -1}{p(0,\vert l\vert -1\vert 0,0)}
f(g)\left(\int_g^{g_{max}}f(x)\, dx\right)^{-1}
\ \ \ \ (g\stackrel{f.b.}{\longrightarrow} g_{max}).
\end{equation}
According to Eq.\ (\ref{eq2.9}), it is easily seen that for all $l\ne\pm 1$, $p(g,\pm 1)/p(g,l)\rightarrow +\infty$ as $g\rightarrow g_{max}$, which proves the concentration of $l$ onto $l=\pm 1$ in this limit, as expected.

Now, it can be interesting to compute the asymptotic behavior\ (\ref{eq2.9}) explicitly for different classes of jump distributions $f(\eta)$. In the following we will do it for $f(\eta)$ algebraic near $\eta =g_{max}$ and $f(\eta)$ with an essential singularity at $\eta =g_{max}$.
%
%
\subsection{Algebraic $\bm{f(\eta)}$ near $\bm{\eta = g_{max}}$}\label{sec2.2}
Consider $f(\eta)$ such that
\begin{equation}\label{eq2.10}
f(\eta)\sim c(g_{max}-\eta)^{\alpha}
\ \ \ \ (\eta\stackrel{f.b.}{\longrightarrow} g_{max}),
\end{equation}
with $c>0$ and $\alpha >  -1$. Note that Eq.\ (\ref{eq2.10}) also includes all the discontinuous $f(\eta)$ at $\eta =g_{max}$, which corresponds to the case $\alpha =0$ with $f(g_{max}^-) =c$ and $f(g_{max}^+)=0$. (For $-1<\alpha<0$ one has $\lim_{\eta\stackrel{f.b.}{\longrightarrow} g_{max}}f(\eta)=+\infty$ and $f(g_{max}^+)=0$ with integrable singularities at $\eta =\pm g_{max}$). From\ (\ref{eq2.9}) and\ (\ref{eq2.10}) one immediately gets
\begin{equation}\label{eq2.11}
\frac{p(g,\pm 1)}{p(g,l)}\sim\frac{(\alpha +1)(\vert l\vert -1)}{p(0,\vert l\vert -1\vert 0,0)}
\, \frac{1}{g_{max}-g}
\ \ \ \ (g\stackrel{f.b.}{\longrightarrow} g_{max}).
\end{equation}
The divergence $\sim (g_{max}-g)^{-1}$ does not depend on $f(\eta)$ and is therefore universal in this class of jump distributions. It holds in particular for a `top-hat' distribution $f(\eta)=(2g_{max})^{-1}\bm{1}_{\lbrace -g_{max}\le\eta\le g_{max}\rbrace}$, which corresponds to $\alpha =0$ and $c=(2g_{max})^{-1}$ in\ (\ref{eq2.10}). Numerical simulations displayed in Sec.\  \ref{sec4} (see Fig. \ref{Fig:concent_RW}) clearly show a good agreement with our predictions (\ref{eq2.11}).
%
%
\subsection{$\bm{f(\eta)}$ with an essential singularity at $\bm{\eta = g_{max}}$}\label{sec2.3}
Consider $f(\eta)$ such that
\begin{equation}\label{eq2.12}
f(\eta)\sim c\exp\left\lbrack -\frac{1}{(g_{max}-\eta)^{\alpha}}\right\rbrack
\ \ \ \ (\eta\stackrel{f.b.}{\longrightarrow} g_{max}),
\end{equation}
with $c>0$ and $\alpha >0$. For $g$ close to $g_{max}$ one has
\begin{eqnarray}\label{eq2.13}
\int_g^{g_{max}}f(x)\, dx&\sim&\frac{c}{\alpha}\int_{(g_{max}-g)^{-\alpha}}^{+\infty}
u^{-1/\alpha -1}{\rm e}^{-u}du \\
&\sim&\frac{c}{\alpha}(g_{max}-g)^{\alpha +1}\exp\left\lbrack -\frac{1}{(g_{max}-g)^{\alpha}}\right\rbrack
\ \ \ \ (g\stackrel{f.b.}{\longrightarrow} g_{max}), \nonumber
\end{eqnarray}
where we have used\ (\ref{eq2.12}) and made the change of variable $u=(g_{max}-x)^{-\alpha}$. From\ (\ref{eq2.9}),\ (\ref{eq2.12}), and\ (\ref{eq2.13}) one gets
\begin{equation}\label{eq2.14}
\frac{p(g,\pm 1)}{p(g,l)}\sim\frac{\alpha (\vert l\vert -1)}{p(0,\vert l\vert -1\vert 0,0)}
\, \frac{1}{(g_{max}-g)^{\alpha +1}}
\ \ \ \ (g\stackrel{f.b.}{\longrightarrow} g_{max}).
\end{equation}
The divergence $\sim (g_{max}-g)^{-\alpha -1}$ now depends on $\alpha$ and is faster than in the previous case ($\alpha >0$), albeit still algebraic.
%
%
\section{Continuous time random walks}\label{sec3}
Let $p(g,t)$ denote the joint PDF of the gap $g$ and the time interval $t$ between the first two maxima of the infinite sequence $\lbrace x_0, x_1, x_2, \cdots\rbrace$ [see Eq.\ (\ref{lim_PDF2})]. From the concentration of $p(g,l)$ onto $l=\pm 1$ and $p(g,t)=\sum_{0<l\le n}p(t\vert l)\, p(g,l)$, one readily has
\begin{equation}\label{eq3.1}
p(g,t)\sim p(t\vert l=1)p(g,l=1)\ \ \ \ (g\stackrel{f.b.}{\longrightarrow} g_{max}),
\end{equation}
at fixed $t$, which reads, using $p(t\vert l=1)=\Psi(t)$ and the first Eq.\ (\ref{eq2.7}),
\begin{equation}\label{eq3.2}
p(g,t)\sim\Psi(t)\, f(g)\int_{g}^{g_{max}}f(x)\, dx\ \ \ \ (g\stackrel{f.b.}{\longrightarrow} g_{max}).
\end{equation}
This asymptotic behavior corresponds to the contribution of realizations where the walker gets stuck for a given time $t$ at the second maximum and then jumps directly to the first maximum, [this is the very meaning of Eq.\ (\ref{eq3.1})]. On the other hand, the probability for the walker to stay frozen for a time $t$ decreases to zero as $t$ gets arbitrarily large and the contribution of other realizations, with several jumps between the first two maxima, may become significant or even dominant if $t$ increases fast enough to $+\infty$ (relative to the speed at which $g_{max}-g$ decreases to $0$). Thus, the asymptotic behavior of $p(g,t)$ for both $g\stackrel{f.b.}{\longrightarrow} g_{max}$ and $t\rightarrow +\infty$ requires a more careful analysis.

We follow the same line as in Sec. V.C.1 of Ref.\ \cite{MMS2016}. Writing $\ln\lbrack 1-s\hat{f}(k)\rbrack$ on the right-hand side of\ (\ref{def_Phi}) as $\ln\lbrack 1-\hat{f}(k)\rbrack +\ln\lbrack 1+(1-s)\hat{F}(k)\rbrack$ with $\hat{F}(k)=\hat{f}(k)/(1-\hat{f}(k))$, one finds that the asymptotic behavior of $\phi(\lambda ,s)$ for both $\lambda\rightarrow +\infty$ and $s\rightarrow 1$ reads
\begin{eqnarray}\label{eq3.3}
&&\phi(\lambda ,s)\sim 1-\frac{1}{\pi\lambda}\int_0^{+\infty}\ln\lbrack 1-\hat{f}(k)\rbrack\, dk \nonumber \\
&&-\frac{\sqrt{1-s}}{a\pi\lambda}\int_0^{+\infty}\ln\left( 1+\frac{1}{\overline{k}^2}\right)\, d\overline{k}\ \ \ \ (\lambda\rightarrow +\infty\ {\rm and}\ s\rightarrow 1),
\end{eqnarray}
where we have made the change of variable $k=(1-s)^{1/2}\overline{k}/a$. Injecting\ (\ref{eq3.3}) into the first line of\ (\ref{eq2.2}) and using the fact that the integral over $\overline{k}$ is equal to $\pi$, one has
\begin{eqnarray}\label{eq3.4}
u(x,s)&\sim&\delta(x-0^+)-\frac{1}{\pi}\int_0^{+\infty}\ln\lbrack 1-\hat{f}(k)\rbrack\, dk \nonumber \\
&-&\frac{\sqrt{1-s}}{a}\ \ \ \ (x\rightarrow 0\ {\rm and}\ s\rightarrow 1),
\end{eqnarray}
leading to
\begin{eqnarray}\label{eq3.5}
&&I_1(g,s)\sim s\left\lbrack f(g)-\frac{1}{\pi}\int_g^{g_{max}}f(x)\, dx\, 
\int_0^{+\infty}\ln\lbrack 1-\hat{f}(k)\rbrack\, dk\right\rbrack \nonumber \\
&&-\frac{\sqrt{1-s}}{a}\int_g^{g_{max}}f(x)\, dx
\ \ \ \ (g\stackrel{f.b.}{\longrightarrow} g_{max}\ {\rm and}\ s\rightarrow 1),
\end{eqnarray}
which reduces to
\begin{equation}\label{eq3.6}
I_1(g,s)\sim sf(g)-\frac{\sqrt{1-s}}{a}\int_g^{g_{max}}f(x)\, dx
\ \ \ \ (g\stackrel{f.b.}{\longrightarrow} g_{max}\ {\rm and}\ s\rightarrow 1),
\end{equation}
since in the limit $g\stackrel{f.b.}{\longrightarrow} g_{max}$ the second term on the first line of\ (\ref{eq3.5}) is negligible compared to the first one. From Eqs.\ (\ref{def_GF}),\ (\ref{eq2.5}),\ (\ref{eq3.6}) and\ (\ref{def_gamma}) one gets
\begin{eqnarray}\label{eq3.7}
\tilde{p}(g,\hat{\Psi}(q))&\sim& \int_g^{g_{max}}f(x)\, dx \\
&\times&\left\lbrack\hat{\Psi}(q)f(g)-\frac{(\tau_c q)^{\gamma/2}}{a}\int_g^{g_{max}}f(x)\, dx\, 
\right\rbrack
\ \ \ \ (g\stackrel{f.b.}{\longrightarrow} g_{max}\ {\rm and}\ q\rightarrow 0). \nonumber
\end{eqnarray}
Inverse Laplace transforming\ (\ref{eq3.7}) with respect to $q$ yields
\begin{eqnarray}\label{eq3.8}
p(g,t)&\sim& \int_g^{g_{max}}f(x)\, dx \\
&\times&\left\lbrack\Psi(t)f(g)+\frac{\mathcal{D}_{\rm II}}{a\tau_c (t/\tau_c)^{1+\gamma/2}}\int_g^{g_{max}}f(x)\, dx\, \right\rbrack
\ \ \ \ (g\stackrel{f.b.}{\longrightarrow} g_{max}\ {\rm and}\ t\rightarrow +\infty), \nonumber
\end{eqnarray}
with
\begin{equation}\label{def_D2}
\mathcal{D}_{\rm II}=\frac{\sin(\pi\gamma/2)\Gamma(1+\gamma/2)}{\pi}.
\end{equation}
Equation\ (\ref{eq3.8}) is the counterpart of Eq. (84) in Ref.\ \cite{MMS2016} for jump distributions with a bounded support. It gives a uniform expression of $p(g,t)$ for $f(\eta)$ with a bounded support when both $g\stackrel{f.b.}{\longrightarrow} g_{max}$ and $t\rightarrow +\infty$, and it reduces to the `one-jump' contribution\ (\ref{eq3.2}) in the limits $g\stackrel{f.b.}{\longrightarrow} g_{max}$ then $t\rightarrow +\infty$, as it should be.

For $0<\gamma <1$, $\Psi(t)$ has an algebraic tail at large $t$, $\Psi(t)\sim [-\Gamma(-\gamma)]^{-1}\tau_c^\gamma t^{-1-\gamma}$, which follows from the small $q$ behavior of $\hat\Psi(q)$, Eq.\ (\ref{def_gamma}), and for an algebraic $f(\eta)$ near $\eta =g_{max}$, this large $t$ behavior of $\Psi(t)$ together with Eqs.\ (\ref{eq2.10}) and\ (\ref{eq3.8}) yield the scaling form
\begin{eqnarray}\label{eq3.9}
p(g,t)\sim
\frac{a c^2}{\tau_c}
\, \frac{(g_{max}-g)^{3+2(\alpha +1/\gamma)}}{\lbrack a(\alpha +1)\rbrack^{3+2/\gamma}}
\, K\left(\frac{g_{max}-g}{a(\alpha +1)}\left(\frac{t}{\tau_c}\right)^{\gamma/2}\right)&& \nonumber \\
(g\stackrel{f.b.}{\longrightarrow} g_{max}\ {\rm and}\ t\rightarrow +\infty),&&
\end{eqnarray}
where $K(y)$ is the same scaling function as in Eq. (85) of Ref.\ \cite{MMS2016}. Namely,
\begin{equation}\label{eq3.10}
K(y)=\frac{1}{y^{1+2/\gamma}}\, \left(\frac{\mathcal{D}_{\rm I}}{y}+\mathcal{D}_{\rm II}\right) ,
\end{equation}
with
\begin{equation}\label{def_D1}
\mathcal{D}_{\rm I}=\frac{\sin(\pi\gamma)\Gamma(1+\gamma)}{\pi},
\end{equation}
and where we have used the reflection formula $\Gamma(-z)\Gamma(z+1)=-\pi/\sin(\pi z)$. It is readily seen that $K(y)$ has the following large and small argument behaviors
\begin{equation}\label{eq3.11}
K(y)\sim\left\lbrace
\begin{array}{ll}
\mathcal{D}_{\rm I} \, y^{-2(1+1/\gamma)}&(y\rightarrow 0), \\
\mathcal{D}_{\rm II} \, y^{-1-2/\gamma}&(y\rightarrow +\infty).
\end{array}\right.
\end{equation}
In the plane $(t,g)$ with $g<g_{max}$, the curve $g\sim g_{max}-a(\alpha +1)(\tau_c/t)^{\gamma/2}$ corresponds to the cross-over from a `concentration' -- or `one-jump' --  regime (for $g_{max}-a(\alpha +1)(\tau_c/t)^{\gamma/2}<g<g_{max}$) where the walker gets stuck for a long time $t$ at the second maximum and then jumps directly to the first maximum, to a `many-jumps' regime (for $g<g_{max}-a(\alpha +1)(\tau_c/t)^{\gamma/2}$) where she/he travels a long walk of total duration $t$ (with many jumps) between the second and the first maxima.
 
 For $\gamma =1$, there is no scaling form such as\ (\ref{eq3.9}) but the uniform expression\ (\ref{eq3.8}) makes it possible to determine the domains in the $(t,g)$ plane corresponding to the `concentration' -- or `one-jump' --  regime and to the `many-jumps' regime, respectively. Taking for instance $\Psi(t)=\tau_c^{-1}\exp(-t/\tau_c)$, an algebraic $f(\eta)$ near $\eta =g_{max}$, and comparing the two terms on the right-hand side of\ (\ref{eq3.8}), one finds that the `one-jump' regime corresponds to the domain
\begin{equation}\label{eq3.12a}
g_{max}-2\sqrt{\pi}\, a(t/\tau_c)^{3/2}\exp(-t/\tau_c)<g<g_{max},
\end{equation}
and the `many-jumps' regime to the complementary domain
\begin{equation}\label{eq3.12b}
g<g_{max}-2\sqrt{\pi}\, a(t/\tau_c)^{3/2}\exp(-t/\tau_c).
\end{equation}

Finally, the large $t$ behavior of $p(g,t)$ at fixed $g<g_{max}$ is dominated by the contribution of realizations with several jumps between the first two maxima (`many-jumps' regime) and one has\ \cite{MMS2016}
\begin{equation}\label{eq3.13}
p(g,t)\sim
\frac{\sin(\pi\gamma /2)\Gamma(1+\gamma /2)}{\pi a\tau_c}
\, \frac{I_2(g)^2}{(t/\tau_c)^{1+\gamma /2}}
\ \ \ \ (t\rightarrow +\infty),
\end{equation}
with $I_2(g)$ given by Eq.\ (\ref{eq2.5}). It can be checked that Eq.\ (\ref{eq3.8}) coincides with Eq.\ (\ref{eq3.13}) in the limits $t\rightarrow +\infty$ then $g\stackrel{f.b.}{\longrightarrow} g_{max}$, as it should be.

\begin{figure}[htbp]
\begin{center}
\includegraphics [width=9cm] {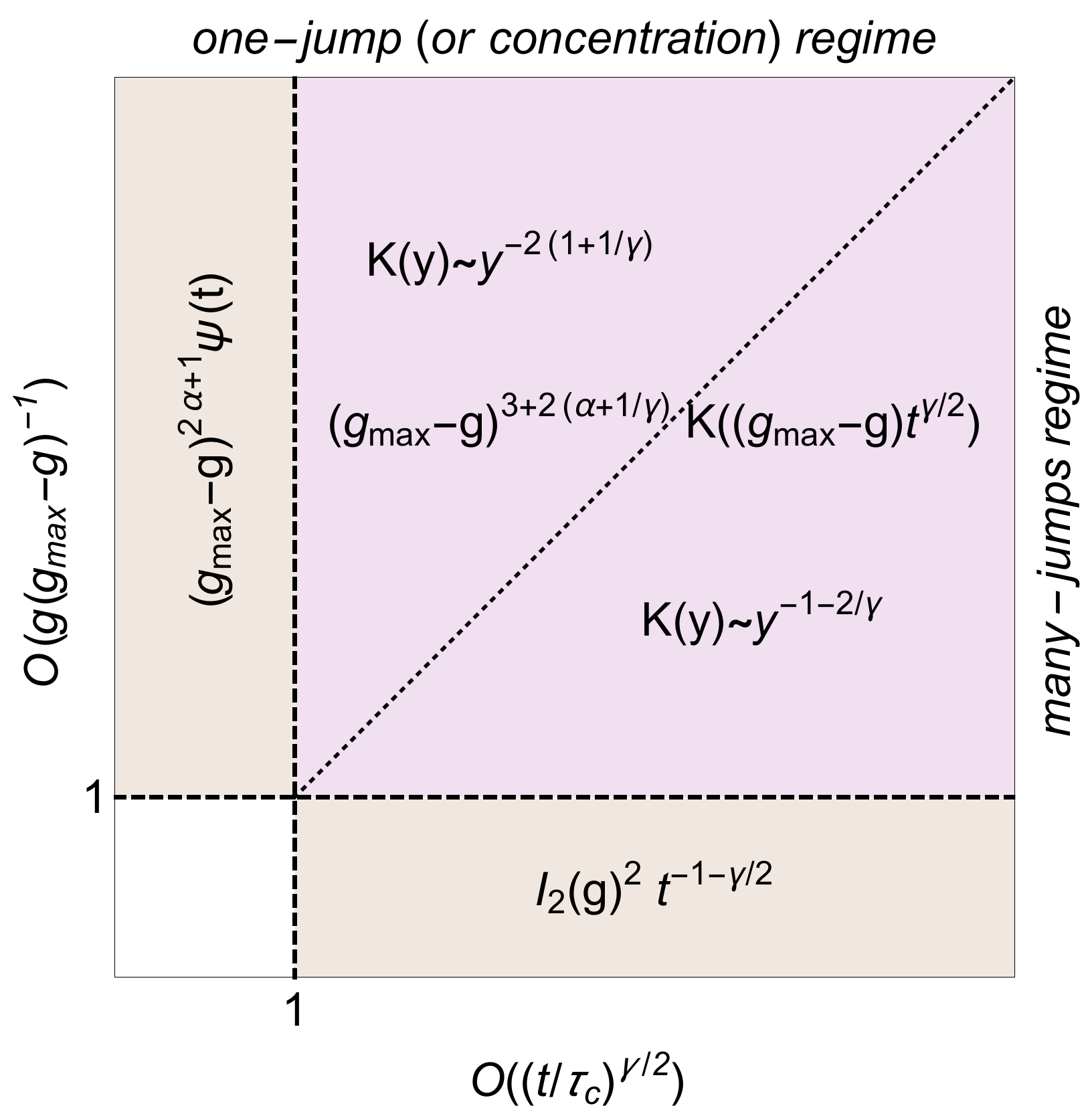}
\caption{\textsl{\it Schematic representation of the asymptotic behaviors of $p(g,t)$ for an algebraic jump distribution near $\eta =g_{max}$ and $0<\gamma <1$ (amplitudes are given in the text and we assume that $a(\alpha +1)/g_{max}=O(1)$). For $t\gg\tau_c$ and $g/(g_{max}-g)\lesssim O(1)$, $p(g,t)$ is given by Eq.\ (\ref{eq3.13}). For both $t\gg\tau_c$ and $g/(g_{max}-g)\sim g_{max}/(g_{max}-g)\gg 1$, one has the scaling form\ (\ref{eq3.9}) with $K(y)$ respectively given by the first line of\ (\ref{eq3.11}) if $(g_{max}-g)t^{\gamma/2}\ll g_{max}\tau_c^{\gamma/2}$ and by the second line of\ (\ref{eq3.11}) if $(g_{max}-g)t^{\gamma/2}\gg g_{max}\tau_c^{\gamma/2}$. For $t\lesssim \tau_c$ and $g_{max}/(g_{max}-g)\gg 1$, $p(g,t)$ is given by Eq.\ (\ref{eq3.2}) with $f(\eta)$ of the form\ (\ref{eq2.10}). The one-jump (or concentration) regime corresponds to the region on the left of the diagonal.}}
\label{fig1}
\end{center}
\end{figure}

Figure\ \ref{fig1} gives a schematic representation of the different asymptotic behaviors of $p(g,t)$ in the plane $\lbrack (t/\tau_c)^{\gamma/2},\, g/(g_{max}-g)\rbrack$ (with $g<g_{max}$) for a jump distribution $f(\eta)$ algebraic near $\eta =g_{max}$ and $0<\gamma <1$. The `one-jump' (resp. `many-jumps') regime is on the left (resp. right) of the diagonal.
\bigskip
%
%
\section{Numerical simulations}\label{sec4}
We have performed numerical simulations to test some of our analytical predictions. For discrete time random walks, we have characterized the effect of concentration, as predicted in Eq. (\ref{eq2.11}), numerically. To this purpose, we have numerically simulated RW of $n$ steps as defined in Eq. (\ref{def_RW}) where the jumps $\eta_i$'s are i.i.d. random variables with jump distribution 
\begin{eqnarray}\label{jump_simu}
f(\eta) = 2^\alpha(\alpha + 1) (1/2 - |\eta|)^{\alpha} \;,  \; -1/2 \leq \eta \leq 1/2 \;,
\end{eqnarray}
which corresponds to the case studied in section \ref{sec2.2}, see Eq. (\ref{eq2.10}) with $g_{max} = 1/2$ and $c= 2^\alpha(\alpha + 1)$. 
Instead of computing the joint PDF $p_n(g,l)$, it is more convenient (numerically) to compute the cumulative distribution function (CDF) $p_n^>(g,l)$ defined by
\begin{eqnarray}\label{def_CDF}
p_n^>(g,l) = \sum_{\ell = l}^n p_n(g,\ell) \;.
\end{eqnarray}
The prediction in Eq. (\ref{eq2.11}) for $p_n(g,l)$ implies that, for fixed $l\ge 2$,
\begin{eqnarray}\label{concent_cumul}
\lim_{n \to \infty} \frac{p^>_n(g,1)}{p^>_n(g, l)} \sim \frac{A(\alpha,l)}{g_{max}-g} \;, \ \ \ \ (\eta\stackrel{f.b.}{\longrightarrow} g_{max}),
\end{eqnarray}
where only the amplitude $A(\alpha,l)$ depends on $\alpha$ and $l$ (see Eq. (\ref{eq2.11})), the divergence $\propto (g_{max}-g)^{-1}$ being independent of these parameters. 
\bigskip
\begin{figure}[htbp]
\begin{center}
\includegraphics [width=0.5\linewidth, angle = -90] {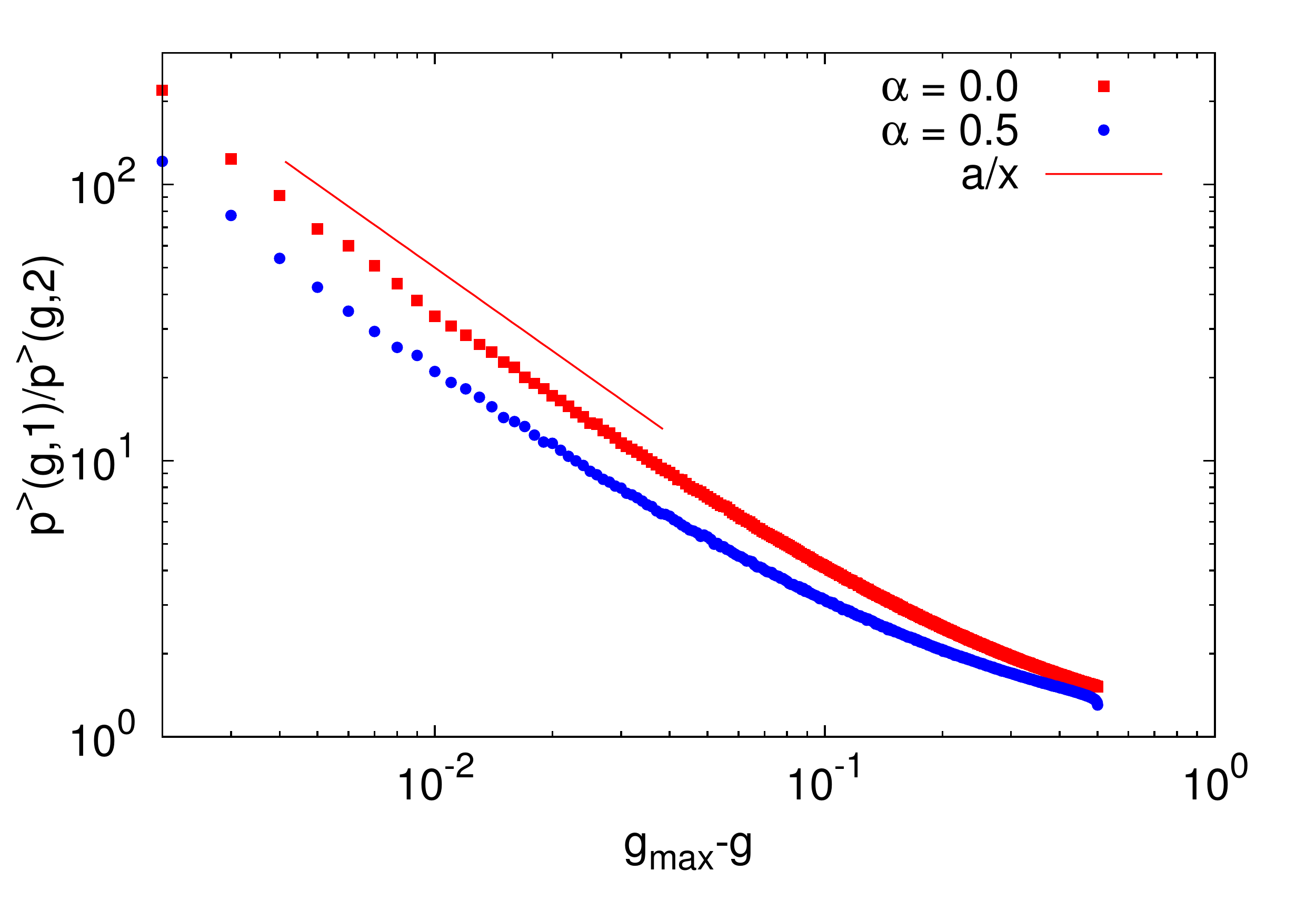}
\caption{\it Plot of $p_n^>(g,1)/p_n^>(g,2)$, as defined in Eq. (\ref{def_CDF}), as a function of $g_{max}-g$ for two jump distributions $f(\eta)$ as given in Eq. (\ref{jump_simu}), for which $g_{max} = 1/2$, for RWs of $n=100$ steps. The different symbols correspond to two different values of $\alpha$: $\alpha = 0$, i.e. the uniform distribution over $[-1/2,1/2]$ (top curve) and $\alpha = 1/2$. The statistics were performed over $10^8$ samples. The solid line is a guide to the eyes, confirming our prediction in Eq. (\ref{concent_cumul}) of the effect of concentration.}\label{Fig:concent_RW}
\end{center}
\end{figure}
In Fig. \ref{Fig:concent_RW} we show a plot of the ratio $p_n^>(g,l = 1)/p_n^>(g,l=2)$ with $n=100$, as a function of $g_{max} - g$ for two different values of $\alpha$. These numerical simulations show a good agreement with our predictions (\ref{concent_cumul}). 

\begin{figure}[hht]
\includegraphics[width = 0.8\linewidth]{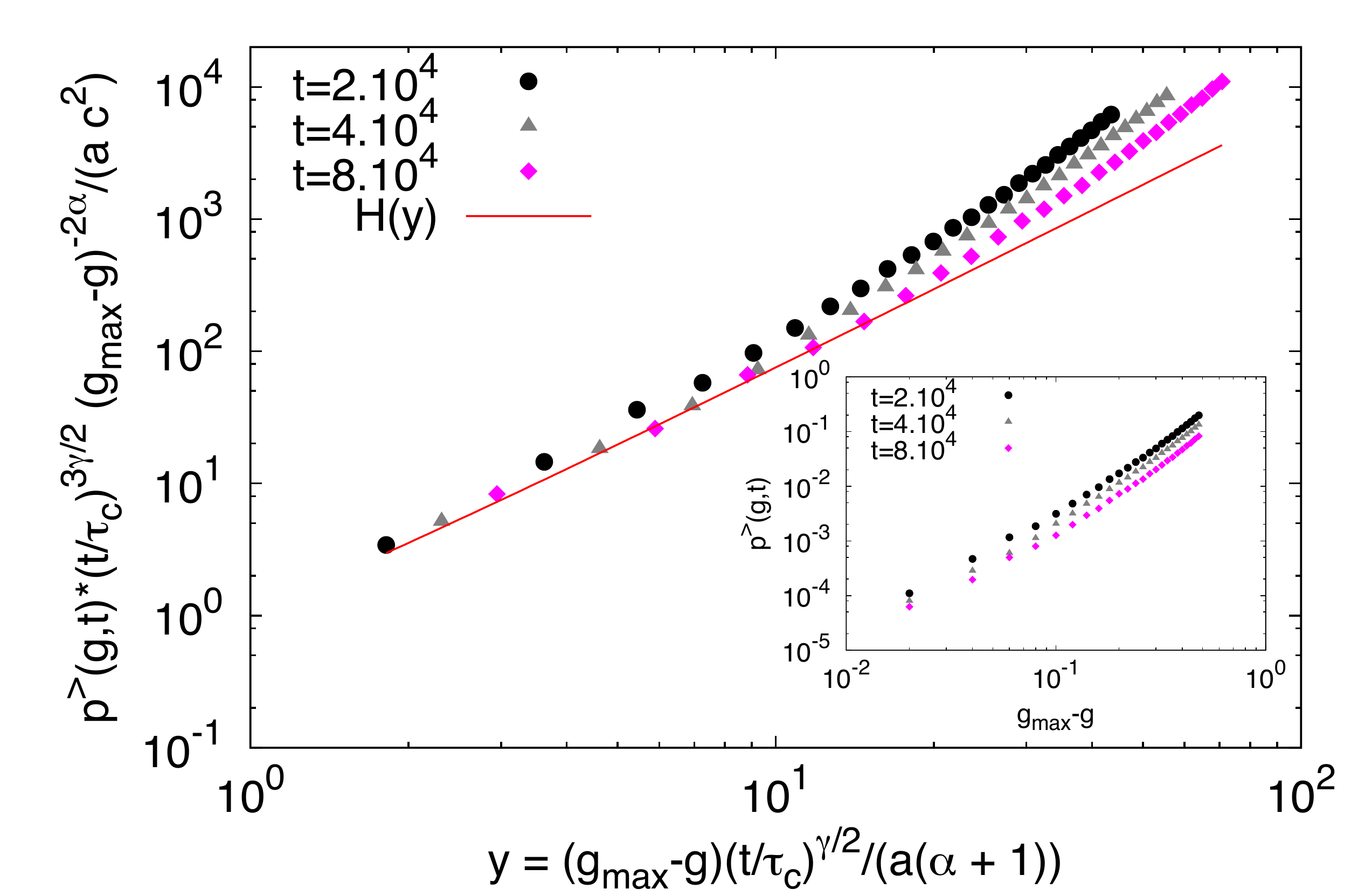}
\caption{\it Scaled plot, in a log-log scale, of $p^>(g,t)(t/\tau_c)^{3\gamma/2} (g_{max}-g)^{2\alpha}/(a\,c^2)$ as a function of $y = (t/\tau_c)^{\gamma/2}(g_{max}-g)/\lbrack a(\alpha +1)\rbrack$, according to the predicted scaling form in Eq. (\ref{scaling_cdf}), for $\alpha = 0$ (and $g_{max} = 1/2$) and $\gamma = 0.7$. The different symbols correspond to different time $t= 2. 10^4, 4. 10^4$ and $t = 8. 10^4$. The statistics were performed over $2. 10^7$ samples. In agreement with our prediction in Eq.~(\ref{scaling_cdf}), this scaled plot indicates that the data for different times $t$ collapse on a single master curve, in the limit $g_{max}-g \ll 1 0$ and $t \gg 1$, which is the solid line, $H(y)$ given in Eq. (\ref{eq:hy}). Note that there is not fitting parameter here. {\bf Inset:} Same data but without any rescaling.}\label{fig:scaling_ctrw} 
\end{figure}

For CTRW, it is much more difficult to observe the effect of concentration on $l = \pm 1$ numerically, 
as this was already noticed in Ref. \cite{MMS2016} in the case of jump distributions defined on the full real line but with a fast decay (typically super-exponential). In the present case, the observation of this regime, corresponding to the first line of Eq. (\ref{eq3.11}), requires to sample jumps extremely close to $g_{max}$ accurately, which is very hard to achieve for long random walks. On the other hand, our numerical simulations in Fig. \ref{fig:scaling_ctrw} show a reasonably good agreement with the scaling form predicted in Eq. (\ref{eq3.9}) in the complementary `many-jumps' regime corresponding to the second line of Eq. (\ref{eq3.11}).

In our simulations, the jumps $\eta$ were drawn from the distribution in Eq. (\ref{jump_simu}) while the waiting times $\tau$'s were drawn from a Pareto distribution of index $\gamma$, as in Ref. \cite{MMS2016}:
\begin{eqnarray}\label{pareto}
\Psi(\tau) = 
\begin{cases}
&0 \;, \;\tau < 1 \\
&\gamma \, \tau^{-\gamma-1} \;, \; \tau > 1 \;.
\end{cases}
\end{eqnarray}
For such Pareto distribution (\ref{pareto}), the characteristic time $\tau_c$ is given by
\begin{eqnarray}\label{tauc}
\tau_c = \left[ \Gamma(1-\gamma)\right]^{1/\gamma} \;.
\end{eqnarray}
Here also, instead of the joint PDF $p(g,t)$, we have computed the CDF $p^>(g,t)$ defined as
\begin{eqnarray}\label{def_cdf}
p^>(g,t) = \int_{t}^\infty p(g,t') \, dt' \;.
\end{eqnarray} 
Our theoretical prediction for the joint PDF in Eq. (\ref{eq3.9}) yields the following scaling form for $p^>(g,t)$:
\begin{eqnarray}\label{scaling_cdf}
p^>(g,t) \sim a\,c^2 \left( \frac{t}{\tau_c}\right)^{-3\gamma/2} (g_{max}-g)^{2\alpha} H\left( \frac{g_{max}-g}{a(\alpha+1)} \left( \frac{t}{\tau_c}\right)^{\gamma/2}\right) \\
{\rm for} \; g\stackrel{f.b.}{\longrightarrow} g_{max}\ {\rm and}\ t\rightarrow +\infty, \nonumber
\end{eqnarray}
where the function $H(y)$ is obtained from the function $K(y)$ in Eq. (\ref{eq3.10}) and is given by
\begin{eqnarray}\label{eq:hy}
H(y) = \frac{{\cal D}_{\rm I}}{\gamma} y + 2 \frac{{\cal D}_{\rm II}}{\gamma} y^2 \;.
\end{eqnarray}
In Fig.~\ref{fig:scaling_ctrw}, we show the results of our numerical simulations where we plot $p^>(g,t)(t/\tau_c)^{3\gamma/2} (g_{max}-g)^{2\alpha}/(a\,c^2)$ as a function of $y = (t/\tau_c)^{\gamma/2}(g_{max}-g)/\lbrack a(\alpha +1)\rbrack$ [see Eq. (\ref{scaling_cdf})] for $\alpha = 0$ and $\gamma = 0.7$. This scaled plot suggest a good collapse of the data for different times $t$ in the regime $g_{max} - g \to 0$ and $t \to \infty$ in agreement with the scaling form predicted in Eq.~(\ref{scaling_cdf}). As discussed above, the small $y$ part of the plot, which corresponds to the concentration (single step) regime, is too rarely sampled to be observed numerically.
%
%
\section{Summary}\label{sec5}
In this paper, we have studied the statistics of the gap $g$ and the time interval between the first two maxima of a RW with bounded jumps, $-g_{max}\le\eta\le g_{max}$. We have considered discrete as well as continuous time RWs in the limit where the number of steps in the walk goes to infinity. In both cases the statistics exhibit an interesting concentration effect by which a gap close to its maximum possible value, $g\approx g_{max}$, is much more likely to be achieved by two successive jumps rather than by a long walk between the first two maxima.

More specifically, for a discrete time random walk, the number of jumps $l$ between the first two maxima is found to concentrate onto $l=\pm 1$ as $g\rightarrow g_{max}$ in the sense that for all $l\ne\pm 1$, $p(g,\pm 1)/p(g,l)\rightarrow +\infty$ as $g\rightarrow g_{max}$. Furthermore, for all jump distributions of practical interest which behave algebraically as $\eta\rightarrow g_{max}$ (or are discontinuous to zero at $\eta = g_{max}$), this divergence turns out to follow a universal behavior $p(g,\pm 1)/p(g,l)\sim (g_{max}-g)^{-1}$ as $g\rightarrow g_{max}$. This result is clearly confirmed by our numerical simulations.

In the case of continuous time random walks, we have performed a thorough analysis of the different asymptotic behaviors of the joint PDF $p(g,t)$ in the plane $(g,t)$, where $t$ is the time between the first two maxima. (Note that by $p(g,-t)=p(g,t)$, we can restrict ourselves to $t>0$ without loss of generality). The main result of this study is the existence of a scaling form for $p(g,t)$ when $0<\gamma <1$ [see Eq.\ (\ref{def_gamma})] and for algebraic (or discontinuous to zero) jump distribution near $\eta =g_{max}$. Namely, we have shown in this case that there is a scaling regime $a(g_{max}-g)^{-1},\ t/\tau_c \gg 1$ with fixed $a^{-1}(g_{max}-g)(t/\tau_c)^{\gamma/2}$, in which $p(g,t)$ takes the scaling form given in Eqs. (\ref{eq3.9})-(\ref{eq3.11}). For $(t/\tau_c)^{\gamma/2}< a(g_{max}-g)^{-1}$, the statistics is in the `concentration' -- or `one-jump' --  regime where the walker get stuck for a long time $t$ at the second maximum and then jumps directly to the first maximum. In the complementary domain, $(t/\tau_c)^{\gamma/2}> a(g_{max}-g)^{-1}$, the statistics is in a `many-jumps' regime where she/he travels a long walk of total duration $t$ (with many jumps) between the second and the first maxima. For $\gamma =1$, or non algebraic jump distribution near $\eta =g_{max}$, there is no scaling form and\ (\ref{eq3.9}) is replaced with the uniform expression\ (\ref{eq3.8}) from which it is also possible to determine the domains in the $(t,g)$ plane corresponding to `one-jump' --  and `many-jumps' regimes.
Our numerical simulations show a reasonably good agreement with our prediction\ (\ref{eq3.9}) despite the lack of a good sampling of the concentration regime in this case, which prevented us from reaching an accuracy as good as in the discrete time setting.

The present work completes our previous results \cite{MMS2013,MMS2014} and \cite{MMS2016} on the joint PDF of the gap between the first two maxima and the time elapsed between them, for discrete and continuous time random walks. In particular, all our results show a very rich behavior of this joint PDF depending on the distribution of the jumps of the random walk. On the other hand, recent works on the statistics of higher order gaps showed that the (marginal) PDF of the $k$-th gap, between the $k$-th and $(k+1)$-th maxima, becomes universal  in the limit of large $k$ for jump distributions with a well defined second moment \cite{order_rw}. In view of this, it would now be quite interesting to study the joint PDF of the $k$-th gap and the time elapsed between the corresponding two maxima, and investigate the question of large $k$ universality of this joint PDF (with respect to the jump distribution, with or without second moment). This remains a challenging issue.
%
%

\acknowledgements

We wish to thank S. N. Majumdar warmly for many inspiring discussions, since the beginning of this project.

\end{document}